\documentclass[%
 reprint,
superscriptaddress,
 amsmath,amssymb,
 aps,
pra,
]{revtex4-2}

\usepackage{graphicx}
\usepackage{dcolumn}
\usepackage{bm}

\begin{document}

\preprint{APS/123-QED}

\title{High speed microcircuit and synthetic biosignal widefield imaging using nitrogen vacancies in diamond}

\author{James L. Webb}
 \email{jaluwe@fysik.dtu.dk}
\affiliation{Center for Macroscopic Quantum States (bigQ), Department of Physics, Technical University of Denmark, 2800 Kgs. Lyngby, Denmark}%
\author{Luca Troise}%
\affiliation{Center for Macroscopic Quantum States (bigQ), Department of Physics, Technical University of Denmark, 2800 Kgs. Lyngby, Denmark}%
\author{Nikolaj W. Hansen}%
\affiliation{Department of Neuroscience, Copenhagen University, 2200 Copenhagen, Denmark}%
\author{Louise F. Frellsen}%
\affiliation{Center for Macroscopic Quantum States (bigQ), Department of Physics, Technical University of Denmark, 2800 Kgs. Lyngby, Denmark}%
\author{Christian Osterkamp}%
\affiliation{Institute for Quantum Optics and Center for Integrated Quantum Science and Technology (IQST), Ulm University, Albert-Einstein-Allee 11, 89081, Ulm, Germany.}%
\author{Fedor Jelezko}%
\affiliation{Institute for Quantum Optics and Center for Integrated Quantum Science and Technology (IQST), Ulm University, Albert-Einstein-Allee 11, 89081, Ulm, Germany.}%
\author{Steffen Jankuhn }%
\affiliation{Division Applied Quantum System, Felix Bloch Institute for Solid State Physics, Leipzig University, 04103, Leipzig, Germany}%
\author{Jan Meijer}%
\affiliation{Division Applied Quantum System, Felix Bloch Institute for Solid State Physics, Leipzig University, 04103, Leipzig, Germany}%
\author{Kirstine Berg-S{\o}rensen}
\affiliation{Department of Health Technology, Technical University of Denmark, 2800 Kgs. Lyngby, Denmark}%
\author{Jean-Fran\c{c}ois Perrier}%
\affiliation{Department of Neuroscience, Copenhagen University, 2200 Copenhagen, Denmark}%
\author{Alexander Huck}
 \email{alexander.huck.dtu.dk}
\affiliation{Center for Macroscopic Quantum States (bigQ), Department of Physics, Technical University of Denmark, 2800 Kgs. Lyngby, Denmark}%
\author{Ulrik Lund Andersen}
 \email{ulrik.andersen@fysik.dtu.dk}
\affiliation{Center for Macroscopic Quantum States (bigQ), Department of Physics, Technical University of Denmark, 2800 Kgs. Lyngby, Denmark}%

\date{\today}

\begin{abstract}
The ability to measure the passage of electrical current with high spatial and temporal resolution is vital for applications ranging from inspection of microscopic electronic circuits to biosensing. Being able to image such signals passively and remotely at the same time is of high importance, to measure without invasive disruption of the system under study or the signal itself. A new approach to achieve this utilises point defects in solid state materials, in particular nitrogen vacancy (NV) centres in diamond. Acting as a high density array of independent sensors, addressable opto-electronically and highly sensitive to factors including temperature and magnetic field, these are ideally suited to microscopic widefield imaging. In this work we demonstrate such imaging of signals from a microscopic lithographically patterned circuit at the micrometer scale. Using a new type of lock-in amplifier camera, we demonstrate sub-millisecond (up to 3500 frames-per-second) spatially resolved recovery of AC and pulsed electrical current signals, without aliasing or undersampling. Finally, we demonstrate as a proof of principle the recovery of synthetic signals replicating the exact form of signals in a biological neural network: the hippocampus of a mouse. 

\end{abstract}

\maketitle

\section{Introduction}

Microscopic electrical transport underpins both synthetic systems such as integrated circuits as well as biological processes  including the functioning of the human brain and nervous system. Operating at high speed and relying on transport down to single electronic charges, these systems require advanced inspection tools, in order to monitor transport performance and diagnose fault. In the case of a synthetic circuit, faults arising from factors including poor insulation and electromigration \citep{HoffmannVogel2017} can reduce device lifetime, requiring intervention before failure. Equivalently in a biological system, faults at the microscopic level arising from disease (e.g. Alzheimer's) can have serious consequences if left undetected. 

Key to microscopic inspection are techniques capable of spatially and temporally resolving electrical transport in both synthetic or biological systems. Presently, this is difficult to achieve noninvasively without causing damage to the circuit or surrounding packaging - a factor especially important for biological tissue. Existing methods for integrated circuit inspection such as laser voltage probing \citep{Kindereit2014}, electron microscopy \citep{Nakamae2021}, THz spectroscopy\citep{True2021} and electrophysiology probes\citep{Ulyanova2019} or fluorescence microscopy \citep{Ohki2005} for biosystems require unimpeded local access. Furthermore, these existing techniques are all active, requiring direct interaction with the target system. This active sensing has the potential to interfere with the target signal or at worst induce damage in the system under study.  

What is desirable is an inspection tool that is passive, remote and noninvasive. In recent years, a new technique has emerged for this purpose utilising point defects in solid state materials. Located in a solid material at a distance from the signal source under study, these can act as atomic scale (quantum) sensors to remotely and passively probe factors including electric field\citep{Dolde2011}, temperature\citep{Neumann2013}, pressure/strain\citep{Knauer2020}, motion \citep{Cohen2020,Liu2020} and in particular magnetic field\citep{Taylor2008, Hong2013, Wolf2015}. State of the art measurements are based on negative charged nitrogen-vacancy centres (NVs) in diamond \citep{Gruber1997}. Consisting of a substitutional nitrogen dopant paired with a lattice vacancy, these defects have an energy level structure highly sensitive to environmental factors. Sensing using NV centres can be performed by monitoring fluorescence output of a single or NV ensemble in a diamond under green laser and microwave irradiation via optically detected magnetic resonance (ODMR) spectroscopy \citep{Delaney2010,Levine2019}. Acting as a high density array of independent sensors, NVs are ideally suited for widefield imaging using fluorescence emission, particularly imaging of magnetic field arising from electrical circuits\citep{Horsley2018, Chen2021}, electronic transport in graphene\citep{Tetienne2017,Ku2020}, ferromagnetic geological samples\citep{Glenn2017} and in biological systems\citep{Price2020,LeSage2013, Schirhagl2014}. 

A particular goal of widefield NV sensing is to image and record the passage of signals in a living biological neural network \citep{Hall2012}. A realisation of this idea would give key new physiological insight. However, state of the art imaging with NV centres suffers from two disadvantages. The first is that the level of per-pixel sensitivity is as-yet insufficient to resolve the picotesla-level signals from neurons\citep{Parashar2020}. This is compounded by limitations in field-of-view as well as artifacts and background noise \citep{Webb2020}. The second is that limitations in measurement methods, in particular low camera frame rates, prohibit the temporal resolution of the desired signals. Imaging has therefore been limited to slow or static measurements (of e.g. temperature \citep{Tanos2020,Chen2021} or DC magnetic field from ferromagnetic materials\citep{Simpson2016,Broadway2020,Kazi2021}) or relying on signal aliasing \citep{Mizuno2020}, unsuitable for accurate time resolved signal recovery. 

In this work, we address the second of these limitations. We use a novel type of lock-in amplifier camera (Helicam, Heliotis AG \citep{Lambelet2011,Wojciechowski2018,Patel2011}), imaging NV centres in diamond to spatially resolve the absolute magnitude of the magnetic field induced by current flow in a lithographically patterned microcircuit. We demonstrate the viability of our technique to map current flow in the circuit with a wide field of view (mm scale) and micrometer-level resolution, limited only by choice of microscope objective. We show this can be done with a simple constant laser and microwave (continuous wave) method and non-uniform laser illumination of the diamond at Brewster's angle. This optical coupling method allows us to maximise coupling of laser pump light into the diamond, key to increasing overall fluorescence emission and hence improving overall sensitivity.  We demonstrate imaging of AC current signals of frequencies up to 1.5kHz and recover pulsed signals replicating digital and analog signals in integrated circuits with sub-millisecond resolution. This is significantly faster than current state of the art experiments in the field. We demonstrate signal recovery above the noise floor at the single pixel level, using image regions of the diamond both directly adjacent to and well away from the patterned wire. Finally, we generate and demonstrate the detection of a signal precisely replicating the shape of current signals (field excitatory postsynaptic potentials, fEPSPs) as measured from the hippocampus in the living brain of a mouse, as a proof of principle demonstration towards biological neural network imaging. 

\section{Methods}

\begin{figure}
\includegraphics[width=8.6cm]{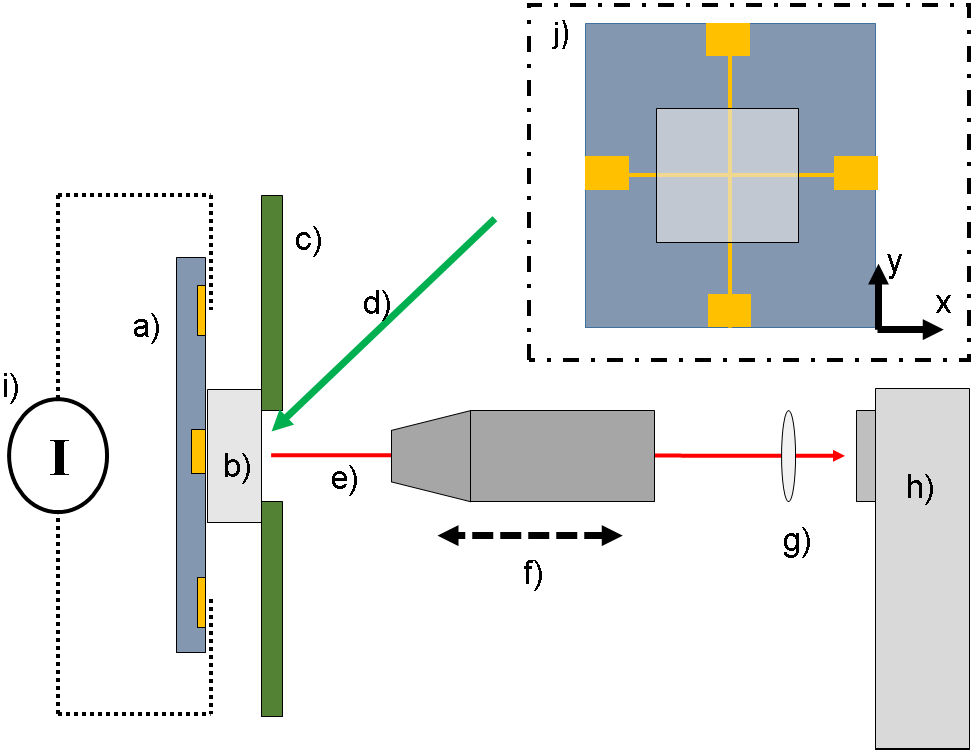}
\caption{Simplified schematic of the experimental setup (not to scale). The lithographically patterned circuit cross on a Si or glass substrate (a) is placed adjacent to the diamond (b) and affixed using either acrylate glue or a 3D printed plastic holder. The diamond and circuit is then affixed to a PCB microwave antenna (c) with a hole to allow pump laser (d) to reach the diamond at Brewster's angle and fluorescence collection (e) via a translatable microscope objective (f). The fluorescence is focused (g) into the lock-in camera (h) to produce an image of the circuit. Current is applied through the circuit (h) using soldered contacts to the circuit chip. The view from the perspective of the camera is detailed in (j), with the diamond on the centre of the cross tracks.  }
    \label{figset}
\end{figure}

A simplified schematic of our setup is shown in Figure \ref{figset}. We use a 2x2mm electronic grade diamond (Element 6) with a top 1µm CVD overgrown layer with approximately 99.99\% $^{12}$C and 10ppm $^{15}$N doping. The diamond was  irradiated with He$^{+}$ at 1.8MeV with a dose of 10$^{15}$ cm$^{-2}$, followed by annealing at 900$^{\circ}$C. We measured the NV concentration to be in the range 0.1-1ppm. The fluorescence emission of this diamond, measured using a photodiode power meter (Thorlabs S120C) in place of the camera was measured to be 152$\mu$W under 1.8W of green pump laser power, a relatively low level as compared to state of the art schemes\citep{Barry2020}.

We fabricated the circuit on either Si or glass substrate, patterned using ultraviolet mask aligned photolithography and metallised by deposition of Ti/Au. Track widths were 10$\mu$m, decreasing to 5$\mu$m in a 15x15$\mu$m square region at the centre of the cross. We mounted the diamond and circuit on a printed circuit board (PCB) microwave antenna with a hole through which laser light could be directed into the diamond, which was mounted directly onto the PCB using Kapton tape. We mounted the diamond with the NV centre layer at the surface adjacent to that where we attached the circuit, using a small amount of acrylate glue or a 3D printed plastic holder. 

Up to 1.8W of green pump light could be supplied to the diamond at Brewster's angle using a diode pumped solid state laser (Cobolt Samba). Light was directed to the diamond via a focusing lens (Thorlabs LB1676), defocused slightly in order to achieve more uniform illumination across a large part of the diamond. Florescence collection was achieved using a 10x microscope objective (Nikon) placed approximately 10mm from the diamond NV layer and mounted on a micrometer translation stage ($\pm$5mm travel) to achieve focus control, with a second coarse adjustment stage to control lateral optical alignment and a screw post for vertical alignment. Light collected by the objective was passed through a long wavelength pass filter (Thorlabs FELH0600) mounted in a beamtube to remove the 532nm pump laser light, leaving only the emitted diamond fluorescence. A second lens (Thorlabs) placed 20cm behind the objective focused this light into the aperture of the camera. 

For fluorescence detection we used a lock-in amplifier camera (Heliotis Helicam C3), running at a rate of up to 3500 frames per second (fps). Although heatsunk to the optical table and capable of 3800fps, we ran the camera at a reduced framerate due to concerns regarding overheating during long period acquisitions. The camera produced a trigger signal which was supplied to a microwave generator (Stanford SG380). This was used to frequency modulate microwaves (at $\pm$4MHz) which were then delivered to the PCB antenna and diamond via a microwave amplifier (Minicircuits ZHL-16W-43+). When on NV microwave resonance, this modulation produced the same frequency modulation of the fluorescence light emitted by the NV centres. This could be detected and demodulated by the camera, following the same basic principle as a standard lock-in detection measurement for each camera pixel, allowing significant rejection of noise, particularly pump laser DC and low frequency technical noise. Our field of view projected onto the 300x300 pixel sensor area (292x280 useable) with a 10x objective was approximately 450x450$\mu$m. Magnification could be increased if required by the simple replacement of the microscope objective and eyepiece lens. 

To maximise sensitivity, we required the maximum change in measured pixel value $p_v$ recorded by the camera. Here we define $p_v$ as the amplitude derived from the in-phase(I) and quadrature (Q) camera-digitised values for each pixel, with a value ranging from 0-1024 (10 bits). Maximum change in $p_v$ could be obtained by measuring at a single microwave frequency $f$ on the point of greatest slope d$p_v$/d$f$ in the ODMR spectrum. To identify this frequency of maximum sensitivity and to enable conversion back into real (Tesla) units of magnetic field, we performed a reference ODMR spectrum. Here we swept microwave frequency $f$=2700-3100MHz, simultaneously imaging using the camera in lock-in mode while adjusting a static magnetic field to split the resonances along the 4 NV crystallographic axes. This field was applied using a 1-inch square neodymium permanent magnet behind the diamond circuit assembly, with the majority of field out of the circuit plane. We then selected the resonance with the maximum resonance frequency shift produced by the static offset field. Magnetic field imaging was then performed in this resonance, at the microwave frequency $f_{max}$ delivering the maximum ODMR slope (d$p_v$/d$f$) averaged over all image pixels. To precisely identify the point of maximum sensitivity, we then performed a high frequency resolution ODMR reference centred on microwave frequency $f_{max}$$\pm$0.5-2MHz. 

This procedure allowed us to either recover the full ODMR for all NV resonances (a time consuming process) or to more simply determine the ODMR slope for each pixel for a single NV axis just at the microwave frequency giving the maximum magnetic field response. Knowing this slope for a given single NV axis allowed recovery of the magnetic field in Tesla units using the conversion factor d$f$/d$B$=28Hz/nT \citep{Cooper2014}.  

Measurements were performed at different modulation frequencies (2.5KHz to 14kHz) and frame rates (650-3500fps). For each measurement we took 500 continuous frames (the maximum memory capacity of the camera), giving either a single intensity value or a timeseries for each camera pixel. Data was then transferred to PC via USB2.0, taking up to 8sec per 500 frame acquisition. Each timeseries could then be examined individually, to show the presence of the desired pulsed signal, or fast Fourier transformed to recover the frequency and magnitude of the AC signal. Either AC magnitude, pulse magnitude or fluorescence intensity could then be plotted to generate an image of the total magnetic field seen by the NV centres across the field of view. Although our technique also allowed recovery of signal phase by extracting pixel I and Q values separately, for the results in this work we avoid this by using a voltage trigger (NI-DAQ 6221) that set a constant (zero) phase at the start of each 500 frame acquisition. Reference images were taken off microwave resonance and with the circuit grounded through a switchbox in order to subtract camera pixel offset values. 

Signals were generated in the cross circuit using a current source (Keithley 6221), supplying leads soldered to 4 pads patterned on the circuit substrate. The current source was used to generate AC square wave signals as well as pulsed and synthetic biosignals using an inbuilt arbitrary waveform generation (AWG) capability. Currents ranging from 1mA up to 100mA were applied to the circuit. By examining the ODMR traces, no visible drift in resonance frequency associated with temperature increase due to resistive heating was observed. 

To provide the hippocampus biosignal to be replicated in our circuit, brain slices were obtained from adult (4-8 weeks) C57BL/6 mice (Janvier, France). Briefly, following Isoflurane anesthesia mice were decapitated and their brains dissected submerged in ice-cold, carbogen (95\% O2/5\% CO2) saturated sucrose substituted artificial cerebrospinal fluid (sACSF) containing (in mM): Sucrose (200), NaHCO$_3$ (25), glucose (11), KCl (3), CaCl$_2$ (0.1), MgCl$_2$ (4), KH$_2$PO$_4$ (1.1), Sodium pyruvate (2), myoinositol (3) and ascorbic acid (0.5). 300$\mu$m thick sagittal brain slices were cut in ice-cold sACSF using a VT1200s Vibratome (Leica, Germany).  Slices were allowed to recover for at least 90 minutes in an interface type holding chamber, continuously bubbled with carbogen, kept at 28$^{\circ}$C and filled with regular ACSF containing (in mM): NaCl (111), NaHCO$_3$ (25), glucose (11), KCl (3), CaCl$_2$ (2.5), MgCl$_2$ (1.3) and KH$_2$PO$_4$ (1.1). Individual slices containing the hippocampus was transferred to a custom made submerged type recording chamber continuously perfused with carbogen-saturated ACSF. Field excitatory post synaptic potentials (fEPSPs) were evoked by stimulating the Schaffer Collaterals at 0.05Hz using 0.05ms current pulses delivered through a twisted Pt/Ir wire electrode, connected to an A365 stimulus isolator (WPI, USA) and placed in the Stratum Radiatum at the border between the CA3-CA1 regions. fEPSPs were recorded by an ACSF filled glass electrode (4-6 MOhm), connected to a CV-7B headstage  (Molecular Devices, USA) and placed in the Stratum Radiatum region of CA1. fEPSP signals were amplified using a 700B amplifier (Molecular Devices, USA) and digitized for recording (NI-DAQ 6221). 

\section{Results}

\subsection{Intensity and ODMR}

\begin{figure}
\includegraphics[width=8.6cm]{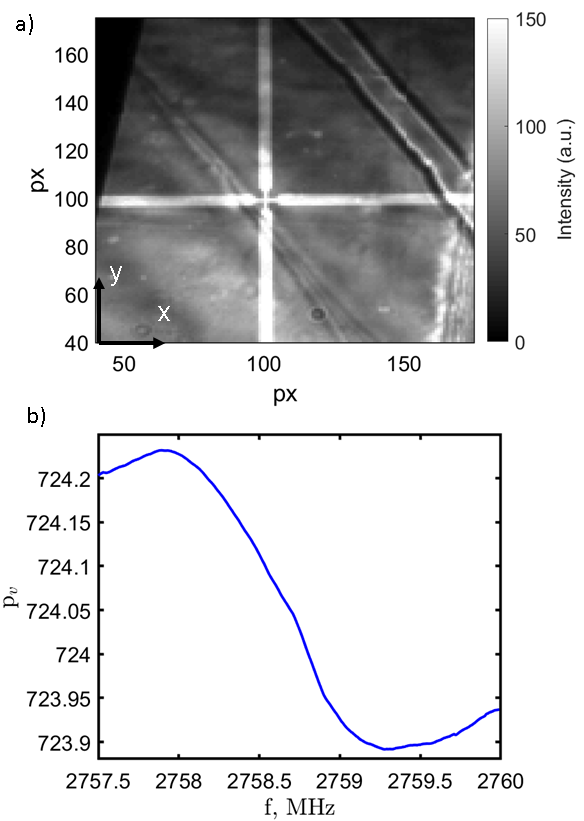}
\caption{a) Image of the intensity of fluorescence from the diamond in a region zoomed in around the cross circuit centre. Throughout this work, 1 pixel $\approx$ t1.5x1.5$\mu$m. The edge of the diamond can be seen as the black area in the upper left of the image. The feature across the upper right is a surface contamination artifact (residual glue on the circuit substrate). b) ODMR spectroscopy as pixel value average across the whole image $p_v$ versus microwave drive frequency centred on a single microwave resonance (single NV axis). This reference trace could be used to convert into real (Tesla) units of magnetic field.  }
    \label{fig1}
\end{figure}

We first aligned and focused the camera and collection optics to give a clear, centred image of the circuit. This was achieved by a simple recording of the NV fluorescence intensity at 50fps while adjusting alignment. Figure \ref{fig1}a,a) shows the circuit imaged through the diamond in intensity mode, with the conductive Ti/Au tracks of the circuit brighter due to reflecting more of the fluorescence generated in the adjacent NV layer. By performing ODMR spectroscopy in intensity mode, we estimated the all-pixel average contrast on microwave resonance to be 1.2-1.6\%.As detailed in Methods, we then used the lock-in capability of the camera, with frequency modulation of the microwaves supplied to the diamond, to perform ODMR spectroscopy. Selecting a single microwave resonance with the strongest response (greatest frequency shift) to the majority out of plane static offset magnetic field, we then performed a detailed step scan (50kHz, 500frames/point) across the frequencies with the maximum ODMR spectrum slope d$p_v$/d$f$ for each pixel. The response averaging all pixels for this spectrum can be seen in Figure \ref{fig1},b). For each pixel, a slightly different ODMR spectrum was recorded due to local broadening effects including strain or variations in offset field, resulting in the less than smooth slope in the figure. Examples of measured single pixel ODMR spectra can be seen in Supplementary Information. The degree of resonance frequency variation was $<$0.2MHz across the image. This allowed us to extract d$p_v$/d$f$ for each pixel using an ODMR scan of 1-2MHz across the resonance, and to remain sensitive (close the maximum slope) for the majority of pixels using only a single fixed microwave frequency ($f$=2758.7MHz).  

We selected a single resonance in this manner due to the significant amount of time required to capture the full ODMR spectrum covering all NV microwave resonances via the relatively slow data transfer rate of the camera. Were faster transfer speeds available, it would be possible to perform vector sensing in the manner similar to that outlined in the work by Schloss et al. \citep{Schloss2018} by recording from each microwave resonance in turn. However, using a single NV axis gave a simple and useful measure of the magnitude of the magnetic field at that point in the image and hence the magnitude of current flow in the adjacent circuit. We note that the ODMR spectrum is only required for magnetic field unit conversion. For the majority of relevant applications often only the relative signal (shape, current path, on/off) is necessary. This can be achieved more simply by taking a fast ODMR trace to find the point of maximum sensitivity at maximum slope d$p_v$/d$f$ and recording the relative response in terms of unitless change in $p_v$.

\subsection{Alternating Current Imaging}

\begin{figure*}
    \includegraphics[width=17.2cm]{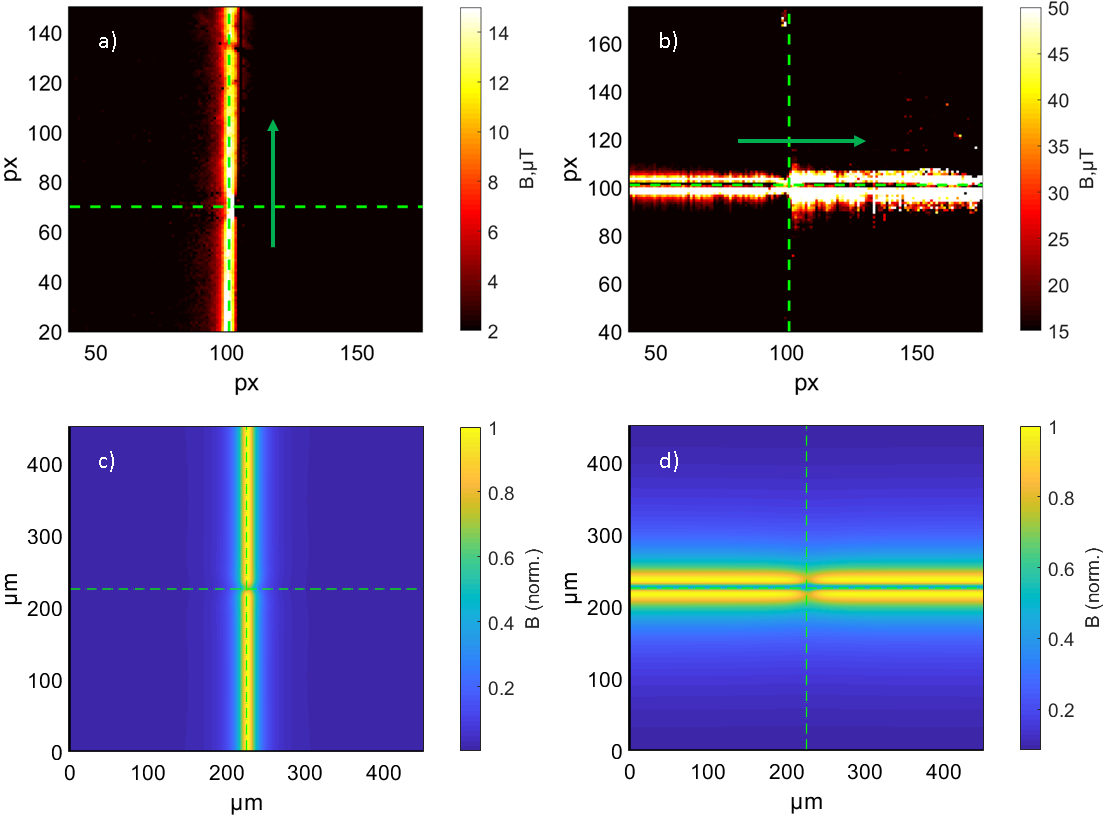}
    \caption{a) Imaged magnetic field resulting from a 130Hz, 4mA alternating current in the vertical (y-axis) cross direction marked with a green arrow, and b) from 130Hz, 20mA current in the horizontal (x-axis) direction. We plot the response only from pixels with a signal SNR$>$3, in order to clearly distinguish the signal from background noise in the images. A strong response was imaged from the NVs directly adjacent to the current-carrying wire tracks, that could be qualitatively replicated by first-principles modeling of the expected field strength (c and d). The location of the cross is shown as a dashed line. The higher noise to the right hand side of the cross in b) as compared to the left is an artifact from higher noise from a block of camera pixels in this region. }
    \label{fig4} 
\end{figure*}

In Figure \ref{fig4},a) and b) we show example images of the magnetic field, arising from a constant 130Hz alternating current passed through the cross circuit either in the vertical (y) or horizontal (x) direction. This current replicates lower frequency current typical of transmission in power lines i.e. 50/60Hz and odd inductive harmonics e.g. 150/180Hz. We use a microwave modulation rate of 2.5kHz, close to the slowest possible camera modulation rate, and framerate of 650fps, delivering the best average per-pixel SNR for the target signal. As expected, we observed the strongest response from the NV centres directly adjacent to the circuit, dropping rapidly away from the wire position. This response was only observed in the directions through the cross where current flowed, indicated in the figures by green arrows. We observed clearly the change in magnetic field resulting from track width reduction in the cross centre, in both current directions. As we imaged a projection of field along a single (out of circuit plane) NV axis, field response was much greater in the vertical (y) current direction, which allowed higher SNR signal recovery at lower current than with current in the horizontal (x) direction. 

In order to validate our images, we calculated the relative strength of the field covering the entire camera field of view (300x300px, approximately 450x450$\mu$m). Modeling as an infinitely long wire of elements $dl$ carrying current $I$ (full details in Supplementary Information) we calculated the field strength projected along an NV axis best aligned with the predominantly out of plane (z direction) DC offset field. Figure \ref{fig4}c),d) shows these plots of the calculated relative magnetic field, normalised to the maximum in the modeled field of view, assuming a circuit-NV separation of 10$\mu$m. The modelled images strongly replicate the experimental images, particularly the central narrowing and strip of relatively low magnetic field near the wire track with current in the horizontal (x) direction, where the magnetic field vector at the NV layer points away from the sensitive NV axis. 

\begin{figure}
\includegraphics[width=8.6cm]{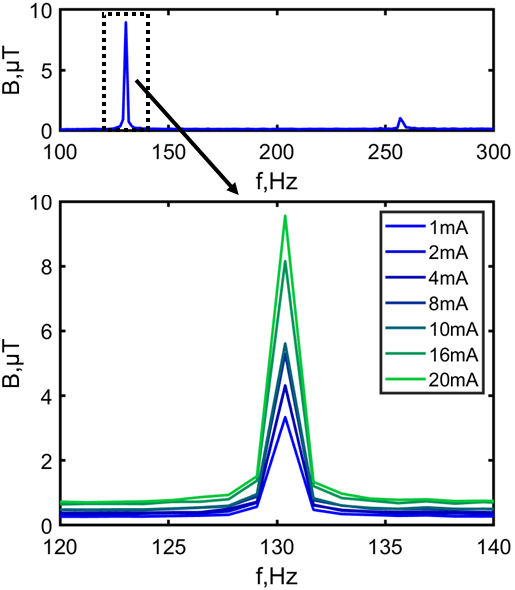}
\caption{Spectrum of the current signal extracted from the FFT of a 500 frame length timeseries, averaging spectra taken from all useable pixels (292x280). Both the 130Hz primary and 260Hz second harmonic are visible. The signal was observed to reduce in strength with lower current, as expected for magnetic field from a current carrying wire.   }
    \label{fig2}
\end{figure}

\begin{figure}
\includegraphics[width=8.6cm]{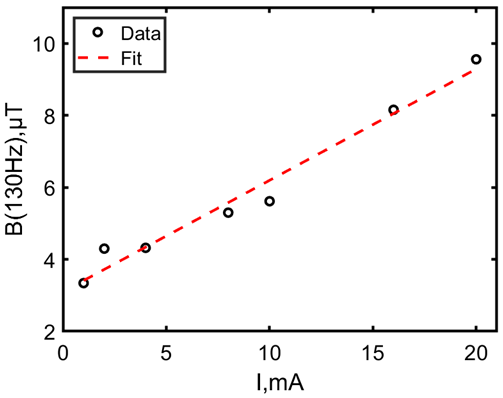}
\caption{Maximum strength of the 130Hz signal from the averaged spectra in Figure \ref{fig2} as a function of current. As expected, the measured magnetic field linearly with current (dashed fit).   }
    \label{fig3}
\end{figure}

To further test magnetic field recording, we imaged the cross at a range of different applied current values between 1-20mA. In Figure \ref{fig2} the amplitude spectrum of the signal can be seen, extracted from a 500 frame timeseries for each pixel then averaged across the image. Examples of the spectra for individual pixels across the image are given in Supplementary Information. The primary (130Hz) and second (260Hz) harmonic of the current signal are both observed and the change in signal amplitude at 130Hz as a function of current can be seen in Figure \ref{fig3}. The AC signal was observed to scale linearly with current, as would be expected from a measurement of magnetic field from a current carrying wire. By increasing the camera modulation rate and framerate, we were able to acquire similar data for AC current signals of frequency up to 490Hz (6kHz microwave modulation, 1000fps) and up to 1.51kHz (14kHz microwave modulation, 3500fps). This can be seen in the Supplementary Information. 


\subsection{Pulsed Current}

\begin{figure}
\includegraphics[width=8.6cm]{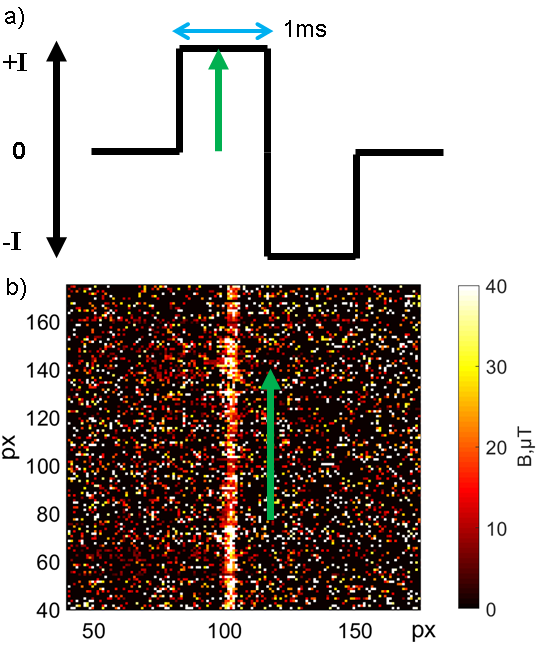}
\caption{a) Schematic of the 1ms/1ms forward/reverse polarity pulsed current signal. The signal is repeated every 20ms. b) Image map of the strength of the 1ms forward component of the pulsed signal extracted from the pixel timeseries over 500 frames (at 3500fps), averaging the 6 pulses captured within the acquisition. For clarity, only pixel values with a signal SNR$>$3 are shown, with the remainder set to zero. The direction of the current through the cross circuit is indicated by a green arrow. }
    \label{fig5}
\end{figure}

\begin{figure}
\includegraphics[width=8.6cm]{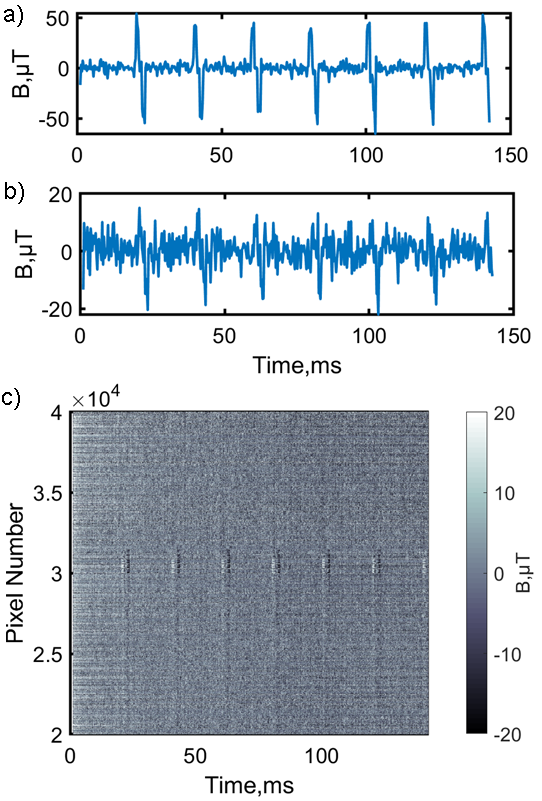}
\caption{Timeseries of the acquired signal for the acquisition period (142ms at 3500fps), taking the average over 1000 repeated  500 frame acquisitions. In a) the average of all pixels with a SNR$>$3 is shown, with the signal clearly resolved. In b) an example timeseries is given for a single pixel in a region of the image  close to the current path, where despite the higher noise level the signal could still be observed. c) A sequence of timeseries for pixels 20000-40000 covering a region centred on the cross current path. The strongest signal is observed for recording from pixels receiving fluorescence from NVs adjacent to the current carrying track, however the pulsed signal is still observable (with a decreasing SNR) away from it.}
    \label{fig6}
\end{figure}

In order to demonstrate detection of a digital signal as typical for an integrated circuit, we applied a series of 20mA current pulses of 1ms/1ms forward/reverse polarity current as detailed in Figure \ref{fig5},a), separated by 20ms intervals. In order to record such short period signals, we increased the camera frame rate to 3500fps and microwave modulation rate to 14kHz, while still capturing 500 image frames, giving a 142ms timeseries for each pixel covering a set of 6 applied pulses. We acquired 2500x500 frame image sets, extracting an average 500 frame timeseries for each pixel. 

The relative strength of the forward current 1ms pulses are shown in the image in Figure \ref{fig5}, b). As a consequence of the higher sensing bandwidth (higher frame rate) we captured more noise, giving a lower image SNR. We also noted a reduction in signal strength, which we attribute to loss of modulation synchronicity between the camera and our microwave generator, arising from trigger incompatibility at modulation rates above approximately 3kHz. Although this technical issue compromised imaged signal strength, we clearly observed the applied pulsed signal in our magnetic field image, directly adjacent to the current carrying track, indicated by a green arrow in Figure \ref{fig5},b). 

The recorded magnetic field from the applied pulses is shown in the timeseries in Figure \ref{fig6}, from a) averaging all pixels with a signal SNR$>$3 and b) for a single example pixel in the image on the circuit track. The 1ms/1ms pulses can be clearly detected, with magnetic field amplitude up to 50$\mu$T. Notably, detection is possible with SNR$>$3 using data from just a single pixel receiving fluorescence from NVs directly adjacent to the circuit track. Although the signal was strongest along the current path, we could still observe the pulses within the images away from this region. This is demonstrated in Figure \ref{fig6},c) plotting the recorded timeseries for 20000 pixels in a region centered on the current track. Although having a lower SNR, the pulsed signal was still observable when imaging NV centre fluorescence up to at a distance of at least 53$\mu$m away from the current path. 

\subsection{Detection of a Synthetic Biosignal}

\begin{figure}
\includegraphics[width=8.6cm]{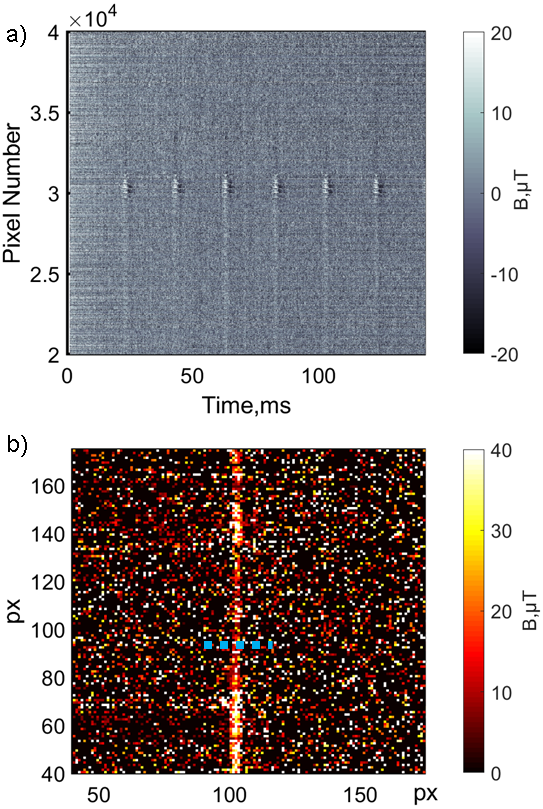}
\caption{The equivalent of plots in Figure \ref{fig5} and Figure \ref{fig6} for the synthetic biosignal. As for the 1ms pulses, the signal can be observed clearly for pixels recording fluorescence from NVs directly adjacent to the current carrying track, but can also be observed at a lower SNR away from this region. The blue dashed line represents the track of pixels for the data presented in Figure \ref{fig7}.}
\label{fig8}
\end{figure}

\begin{figure}
\includegraphics[width=8.6cm]{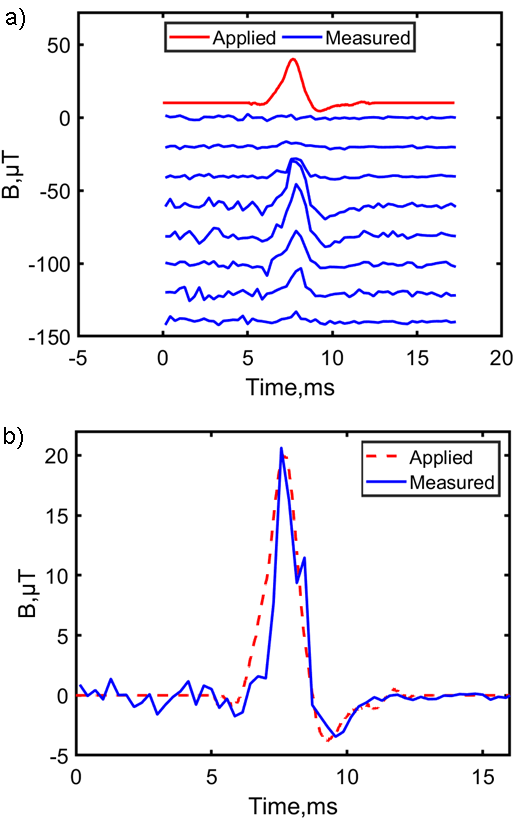}
\caption{a) Magnetic field signal measured from individual pixels taking 5 pixel steps along the blue dashed line indicated in Figure \ref{fig8}. For each step we add an offset from the previous by -20$\mu$T for clarity, with the (scaled) applied synthetic biosignal shown in red at the top. As expected is most clearly observed in the image region closest to the current path, dropping in strength to either side. b) The signal recorded from averaging the response of all pixels with a signal SNR$>$3, overlaid onto the (scaled) generated signal (red, dashed). The signal recovered via the magnetic field imaging matches well with the applied signal, with no distortion or artifacts.}
\label{fig7}
\end{figure}

\begin{figure}
\includegraphics[width=8.6cm]{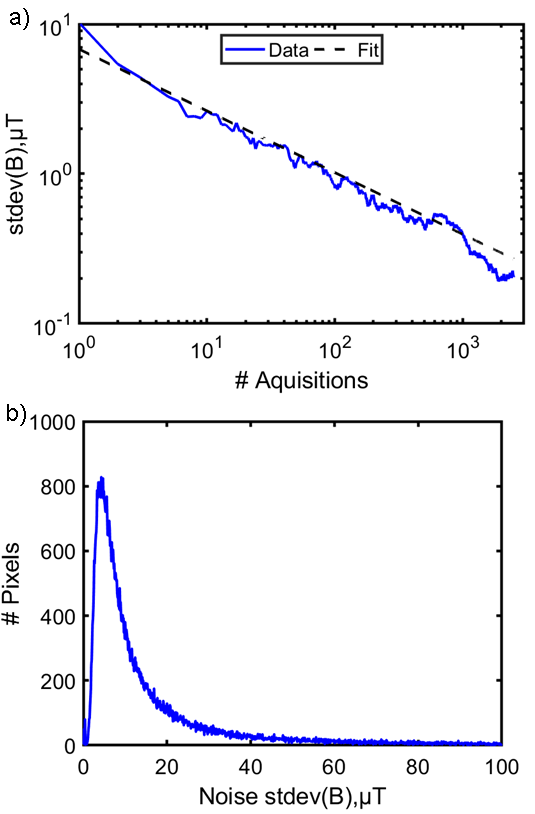}
\caption{a) Scaling of the standard deviation of the magnetic field signal between the current pulses as a function of number of 500 frame acquisitions N=0-2500, as a measure of readout noise. The noise continually reduces with a $\sqrt{N}$ trend, without reaching a plateau. b) Histogram of the individual pixel noise after 1000 acquisitions, with the majority of pixels in the low-$\mu$T noise range.}
\label{fig9}
\end{figure}

Finally, we demonstrate the acquisition of a signal with the same form as a neuronal signal, as may be found in the brain or nervous system of a living person or animal. We replicated the shape of a typical signal acquired from a prior electrical recordings of fEPSPs in the hippocampus of a mouse obtained from living dissected  tissue slices. We applied the signal using the arbitrary waveform generator mode of our current source, with a maximum forward current amplitude of 20mA. We note that this current is significantly more than the current generated by a voltage signal in a neuron (low-nA to pA\citep{Mitterdorfer2002}). However, in this work we purely seek to demonstrate a proof of principle demonstration of lock-in camera acquisition recovery of a signal the same shape and length, for which our choice of amplitude gives a clear visible signal within the sensitivity limitations imposed by our diamond. 

The strength of the detected signal is demonstrated in Figure \ref{fig8},a) and b), mirroring the equivalent Figure \ref{fig5} and \ref{fig6} for the pulsed current, with the strongest signal again imaged closest to the current carrying track. We extracted the synthetic hippocampus signal from averaging the timeseries from 2500x500 frame acquisitions for each pixel. A plot of this signal as compared to the applied current signal can be seen in Figure \ref{fig7},a) for 5 pixel steps across the current track (blue dashed line in Figure \ref{fig8},b). In pixels recording fluorescence from NVs adjacent to the current track we observed a strong signal, highly representative of the applied synthetic signal at a peak magnetic field strength of 15-20$\mu$T, with the signal decaying in strength in the image away from this path. In Figure \ref{fig8},b) we plot the average signal acquired for all pixels with a signal SNR$>$3. This signal also replicated the applied current signal, as shown in the red dashed trace. 

A key feature we note in all our data is the constant reduction in noise with continual averaging (scaling as $\sqrt{N}$, with N the number of 500 frame acquisitions). This is demonstrated in Figure \ref{fig9},a), where we calculate the all-pixel average standard deviation of the magnetic field in the timeseries, taken with zero applied signal in a 5ms period between the current pulses. After 2500 acquisitions, we reach a noise level of approximately 200nT (6.8nT/$\sqrt{Hz}$ on a 1.75kHz acquisition bandwidth at 3500fps), averaging over all 292x280useable pixels in the image. A histogram of the noise on the individual pixels after 1000 acquisitions is given in Figure \ref{fig9},b), peaking at approximately 2.5-3$\mu$T (84-100nT/$\sqrt{Hz}$). The noise follows a Poisson distribution, as would be expected due to fluorescence shot noise with a non-uniform laser illumination of the diamond. In previous experiments using a conventional camera where DC noise is included in the images, we found the noise level to plateau well before the number of frames acquired by the lock-in camera\citep{Webb2020}. We attribute this inclusion of a high level of laser technical noise, in particular fluctuations in laser power. The lock-in camera thus offers the possibility to reach far lower noise levels through continual averaging, with available time the only limitation. The scaling we observe in Figure \ref{fig9} is the same as for our experiments using a DC-noise rejecting balanced photodetector\citep{Webb2021}. 

We note that it was necessary to apply a far larger synthetic biosignal current than would be realistically produced, allowing us to demonstrate in principle the signal acquisition and imaging, but without the necessary noise level to resolve the real signal. As yet, state of the art performance in NV sensing can only resolve real biosignals on the pT-nT level through integrating total fluorescence emission from a mm-scale diamond. Previous theoretical calculations such as the works by Karadas et al. \citep{Karadas2018,Karadas2021} and Wojciechowski et al. \citep{Wojciechowski2018} have explored the expected current and magnetic field strength and camera requirements to image a real biosignal, which have yet to be reached experimentally, as well as the absolute necessity of spatial resolution to avoid cancellation from regions of opposing field polarity.  

The diamond we used in this work has characteristics well below these requirements, with low fluorescence emission and relatively low ODMR contrast (1.2-1.6\%). This is significantly worse than state of the art diamonds in the literature, where total emissions in the mW range at comparable (1-2W) pump laser power is possible and where contrast of up to 30\% can be reached in preferentially aligned samples\citep{Osterkamp2019,Balasubramanian2019,Pham2012}. To generate sufficient contrast, it was also necessary to supply a relatively high microwave power (-5dBm input with 45dB amplifier gain), eliminating the $^{15}$N hyperfine resonances in the ODMR spectrum through power broadening\citep{Drau2011}, further reducing the maximum achievable d$p_v$/d$f$ slope and hence sensitivity. 

We highlight that although we do not have the diamond to reach the necessary levels of sensitivity for real biosensing in this work, no aspect of the measurement method of the synthetic replica signal utilising the lock-in camera is incompatible with future material improvement. In particular, the level of fluorescence from the diamond collected by the camera (a maximum of 152$\mu$W full sensor or 1.7nW/pixel) and the level of the demodulated signal are three orders of magnitude below the lock-in camera limits of up to 2.5-3$\mu$W/pixel or 250-270mW across the sensor area \citep{JMS,Wojciechowski2018}. 

\section{Conclusion}

In this work we demonstrate proof of principle passive and remote imaging of propagating electrical current in a circuit, using a novel type of lock-in amplifier camera. We image the current-induced magnetic field detected via variations in fluorescence emission from nitrogen vacancy (NV) centres in diamond. Using a simple continuous wave method without aliasing or undersampling, we demonstrate imaging of low frequency alternating current (as typical in electrical power distribution systems), rapid pulsed current (as typical in integrated circuits) and a representation of a biosignal (as generated in a neural network in the living brain). We show that each of these types of signal can be observed and mapped, both spatially and as a function of time, at a high frame rate (up to 3500fps) significantly faster than previously achieved in the literature. 

We consider the reduction in noise with continual averaging to be a significant lock-in camera advantage over conventional cameras, where DC and correlated noise from in particular pump laser power fluctuations can cause the noise to plateau after a relatively low number of averaged frames. We consider this feature to be particularly useful in either mapping repeated, consistent fast signals such as transport or magnetism in 2D materials \citep{Broadway2020} or biological neural networks or for slowly varying long period measurements e.g. temperature measurement in living cells via nanodiamond experiments \citep{Fujiwara2020}. 

The method we demonstrate can be readily adapted to a typical inverted microscope setup used for NV sensing and imaging, including for biological samples\citep{Webb2021, Barry2016}. Although our diamond does not allow us to reach the necessary level of sensitivity to directly observe and image signals in a real biosample, with improvements in material growth and optimisation and the potential for implementation of pulsed laser and microwave protocols (recently demonstrated during the production of this work by Hart et al. \citep{Hart2021}), the methods we demonstrate here represent a clear means by which such signals could be resolved in future.

During the preparation of our manuscript, we became aware of contemporary work by Parashar et al. \citep{parashar2021lockin}, independently following a similar procedure as we outline in this work, resolving AC signals from a bulk magnetic field applied across the diamond using a field coil rather than a microcircuit. Both their work and ours highlight the key advantages of the simplicity of the lock-in camera based technique and the limitations, particularly in terms of the possibility of significantly higher sensitivity through improved diamond growth and irradiation. 

\section{Acknowledgments}

The work presented here was funded by the Novo Nordisk foundation through the synergy grant bioQ and the bigQ Center funded by the Danish National Research Foundation (DNRF). We acknowledge the assistance of Kristian Hagsted Rasmussen for microfabrication.

\bibliographystyle{apsrev}
\bibliography{fastrefs.bib}

\end{document}